# RESONANCE CONTROL COOLING SYSTEM FOR A PROTO-TYPE COUPLED CAVITY LINAC[1]

C.A. Treml[2], S.K. Brown, J.D. Bernardin
LANL, Los Alamos NM, 87545, USA

## Abstract

Los Alamos National Laboratory (LANL) is designing a coupled cavity linac (CCL) for the Spallation Neutron Source (SNS) being built at Oak Ridge National Laboratory (ORNL). As part of the design process, a proto-type, Hot Model, consisting of two segments of the SNS CCL was manufactured and installed at LANL. An RF source was applied to the Hot Model to determine the success of the cavity design, the effectiveness of the control algorithms and cooling systems, and the validity of the Resonance Control Cooling System (RCCS) model used during the Hot Model design. From a system controls perspective, an overview of the hot model, a description of the RCCS, and the pertinent data from the RF tests will be presented.

## 1 THE CCL HOT MODEL

The Hot Model is an experimental proto-type of the first two segments of the SNS CCL, and its associated subsystems. These sub systems include water cooling and control, vacuum, RF, and hardware system control. The water cooled components of the CCL are the copper cavities, the short coupling cells between cavities (CCC), and one long coupling cell (LCC) between the two CCL segments. Each of the segments contains eight cavities with their corresponding CCCs.

The desired resonant frequency of the CCL cavities is 805.0 MHz. Under normal operation approximately 70% of the RF Power is dissipated in the CCL cavity walls. For a CCL, electromagnetic field resonant frequency is primarily a function of the geometry of the cavities, side coupling cells, and bridge couplers. As RF power is dissipated in the cavity wall, the copper wall will heat, then expand, and it's resonant frequency will decrease. This frequency shift will be controlled by introducing an opposing frequency shift in the cavity design and a closed loop cavity cooling system.

The CCL cavities have been designed and will be manufactured and pre-tuned to have a resonant frequency of 805.140 MHz with 20°C coolant flowing through the structure surrounding the RF cavity and no RF heating. On the back of the copper structure from which the CCL cavity is machined, there will be four parallel annular paths around the cavity nose (see Fig. 1). These paths will be fed and collected by internal plenums. The internal plenums will be connected to supply and return manifolds that are in turn connected to a water supply, cooling, and control system. Preliminary analysis indicated that for expected operating conditions, i.e. nominal steady RF load, a cooling water flow rate of 2.4 gpm, and an initial water temperature of 20°C, the frequency shift for a high energy ($\beta=0.54$) CCL cavity is a approximately -140 kHz and the temperature rise in the water is approximately 3.0°C. These results prompted the design frequency of 805.140 MHz. The water temperature rise and other deviations in frequency due to errors in manufacturing and power fluctuations are compensated for by changing the water temperature via the RCCS.

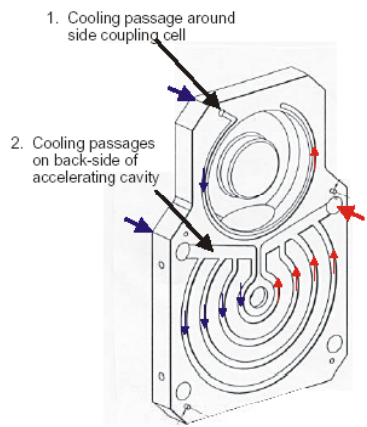

Figure 1: Diagram of the water cooling passages in the CCL RF structure

## 2 RESONANCE CONTROL

The RCCS removes waste heat from the copper structure around the CCL RF cavities and maintains resonance within these cavities through active temperature control. The importance of maintaining the CCL resonant frequency is that it provides for the efficient acceleration of the proton beam with minimal klystron energy. The RCCS is comprised of multiple, closed-loop water cooling systems and the hardware/control for those systems.

---

[1] Work supported by US Department of Energy Contract W-7405-ENG-36
[2] email: treml@lanl.gov

## 2.1 Hot Model/RCCS Layout

The water cooling system for the CCL consists of four main components; the RF structure, transition plumbing, a water skid, and a facility chilled water source. A generalized schematic for the RCCS configuration is shown in Figure 2. The cooling features of the Hot Model RF structure consist of a three main paths; one for each of the cavity coupling cells, one for each of the cavities, and one for the long coupling cell. The CCC path enters on the side of the coupling cell from the inlet manifold and flows around the circumference of the cell and exits on the side of the outlet manifold.. Each of these paths removes approximately 500W/cell of waste heat. Similarly, the cavity path enters then branches into four parallel annular paths on the back side of the RF cavity. This path removes the bulk of the waste heat (~1.87kW/cavity) and is the conduit for the temperature, thus resonance control effort. The water skid is a self-contained unit with all of the necessary plumbing, water treatment hardware, and instrumentation/control features required for delivering water at a desired temperature and constant flow rate to the CCL RF structure. [1]

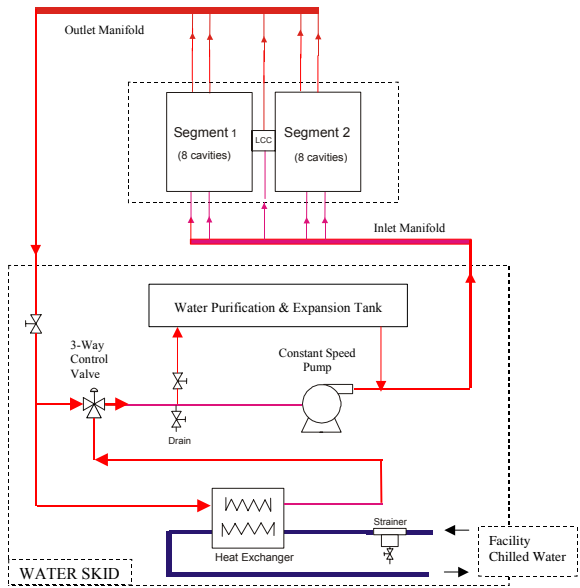

Figure 2: Generalized CCL Cooling System Schematic

## 2.2 Control Scheme

The RCCS control scheme is to maintain the water temperature in the flow loop supplied to the RF structure such that resonance can be maintained in the RF cavities. The water temperature in the flow loop is manipulated by adjusting the proportion of hot water returning from the CCL that is diverted through the water skid heat exchanger. The mechanism for this manipulation is a "3-Way" control valve, whose control effort is calculated via a multi-mode Proportional, Integral, Derivative (PID) control algorithm. All of the instrumentation/control for the Hot Model was implemented via a PC system running *LabVIEW®* and integrated to the hardware/sensors using *Field Point®*.

For the Hot Model, the multimode PID algorithm consisted of manual, temperature feedback (TFB), and RF error feedback (RFEFB) modes. The manual mode allowed an operator to position the control valve by entering values manually into a *LabVIEW®* screen. Manual mode served as a startup, maintenance, and override feature, rather than a mode for normal operation. The RFEFB mode was intended for normal operation, to control resonance under RF Power fluctuations. The TFB mode was used during the loss of RF power, and to preheat or maintain the cavities temperatures close to their resonant temperatures during RF loss.

The RF error is a measure of how far off the cavity RF is from resonance. The RF error signal is calculated using a moving average of the reflected RF power. Whenever the reflected power is measured at zero, i.e. the cavity is in resonance or there is no power, the RF error signal is 5.9V. Hence, 5.9V is the set point for the PID feedback loop when in RFEFB mode. A dead band of oscillations about the resonance frequency of 805.0 MHz (5.9V) of +/-10kHz (+/-0.3V) is specified for the SNS CCL.[2]

In the event of RF power failure or trip, hence, the loss of heat load to the water cooling system, the control mode must be switched from RFEFB to TFB mode. This transition is necessary, since the RF error signal is forced to the set point in the event RF power is lost. With the process variable at the set point, all control effort goes to zero regardless of the temperature of the cavities. In the case where RF power has been lost, the control valve would remain at its current position, and the temperature of the RF cavity would begin to decrease due to the loss of the RF heat load. The motivation for switching to TFB mode is to keep the temperature of the cavities as close as possible to the temperature for which its resonant dimensions are achieved. Hence, there is little time lost returning to resonance once RF power is restored.

While in TFB mode, the desired temperature to use as the process variable would have been the actual cavity temperature; however, it was not possible to place a sensor directly on the cavity wall. Several surface temperature sensors were placed on the outside of the RF structure and water temperature sensors were placed in various locations throughout the water cooling system. Given the sensor information available, the desired process variable for the TFB mode was the water temperature on the transition line leading from the four parallel annular paths on the back side of the RF cavity to the output manifold. This parameter was the most representative of cavity temperature, and had the smallest time delay between it and the cavity temperature changes.

*2.3 Results*

Overall, the RCCS for the Hot Model worked well. The water skid, water transfer lines, orifice plates, manifolds, and the water cooling pathways distributed water evenly and as expected. Both the TFB and RFEFB RCCS control modes brought the process variable into an acceptable region about the set point with settling times on the order of 3 - 5 minutes.

In particular, the results in Figure 3, show the success of the RFEFB mode. This plot shows the RF error tracking the set point of 5.9V over a series of RF power fluctuations, severe drops, and total losses. The RF error tracks inside the dead band within 3 or 4 minutes. A slight drift from the dead band does occur. On analysis of the related data, the cause of this drift was determined to be fluctuations in the facility chilled water. The oscillations of approximately 0.3V about the set point resulted from the coarseness of resolution in the 3-Way control valve. Both the facility chilled water and valve resolution findings will be applied to the SNS CCL.

## 3 MODEL VALIDATION

As part of the RCCS design/tuning process, the data collected from the Hot Model experiments will be used in a model validation scheme to validate the RCCS simulation used during the design of the RCCS. The motivation is to use the validated RCCS model to aid in tuning the RCCS PID control loops for the SNS project.

The validation technique that will be used is being developed as part of the authors dissertation research. Within a model reference adaptive control (MRAC) framework, the nominal model is considered the plant to be controlled or adapted. The model reference, i.e. the desired output of the model is for a specific mode of operation and is inferred using a Bayesian belief network (BBN) utilizing the relationships between the mode or operating conditions of a desired run and experimental data from runs with similar mode or operating conditions. The control logic is provided by a model parameter updating BBN augmented with cost and decision nodes. The updater BBN utilizes expert knowledge of model behavior and statistical knowledge of model input/output relations. The BBN/MRAC scheme minimizes model parameter uncertainty by reducing the difference between the simulation model outputs and the operational run specific reference model outcomes. [3]

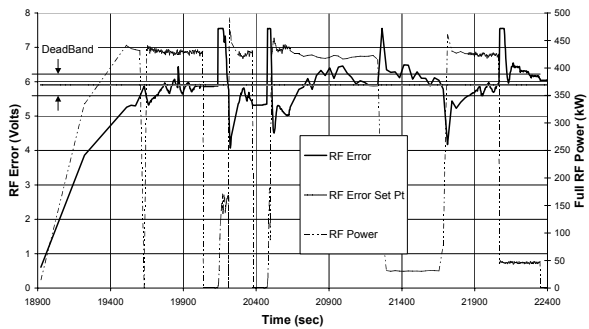

Figure 3: Hot Model Data: RF Error, Set Pt., Dead Band, and RF Power (08/16/01 PM)